\documentclass[oldversion]{aa}
\usepackage{graphicx}

\usepackage{hyperref}

\usepackage{txfonts}

\usepackage{natbib}

\begin{document}

\title{Simultaneous eROSITA and TESS observations of the ultra-active star AB~Doradus}

\author{J.H.M.M. Schmitt\inst{1}\and P. Ioannidis\inst{1}\and J. Robrade\inst{1}\and P. Predehl\inst{2}\and
S. Czesla\inst{1}\and P.C. Schneider\inst{1}}

\institute{Hamburger Sternwarte,
% INST 1
Gojenbergsweg 112,D-21029 Hamburg, Germany\and
Max-Planck-Institut f\"ur extraterrestrische Physik,
85748 Garching, Giessenbachstra\ss e, Germany
}

%\offprints{J-H.M.M. Schmitt }

\mail{J.H.M.M. Schmitt, jschmitt@hs.uni-hamburg.de}
\titlerunning{Superflares on AB Dor}
\date{Received / Accepted }

\abstract{
We present simultaneous multiwavelength observations of the ultra-active star AB~Doradus
obtained in the X-ray range with the eROSITA instrument on board the
 Russian--German Spectrum-Roentgen-Gamma mission (SRG), and in the optical range obtained with
the Transiting Exoplanet Survey Satellite (TESS).   Thanks to its
fortuitous location in the vicinity of the southern ecliptic pole, AB~Dor was observed by these missions
simultaneously for almost 20 days.  With the hitherto obtained data we study the long-term evolution
of the X-ray flux from AB~Dor and the relation between this observable and the photospheric activity of its spots.   Over the 1.5 years of
eROSITA survey observations, the ``quiescent'' X-ray flux of AB~Dor has not changed,  and furthermore it appears unrelated to
the photospheric modulations observed by TESS.  During the simultaneous
eROSITA and TESS coverage, an extremely large flare event with a total energy release of at least 4 $\times$ 10$^{36}$ erg
in the optical was observed, the largest ever seen on AB~Dor.  We show that the total X-ray output of this flare was 
far smaller than this, and discuss whether this maybe a general feature of flares on late-type stars. 
} 


\maketitle
\section{Introduction}

Stellar magnetic activity manifests itself as a variety of different phenomena in different parts of
the outer stellar atmosphere: star spots are located in the photosphere, plage regions in the
chromosphere, and hot, magnetically confined plasma in the corona.   The characteristic
radiation from these different regions is emitted in the optical range, and at UV and X-ray wavelengths.
As the various atmospheric layers are physically interconnected, one needs simultaneous multiwavelength
observations ---preferably over some extended period of time---
to arrive at a better understanding of the interconnection between various activity phenomena.
 
With the advent of the eROSITA and TESS missions (see a detailed discussion in Sect.~\ref{sec_eros_tess}) 
a new observational opportunity has opened up that allows us to monitor stars continuously from space at optical and
X-ray wavelengths.  While the detailed scientific goals and observing strategies for eROSITA and
TESS are quite different, they allow simultaneous coverage over several weeks in smaller regions
around the poles of the ecliptic.  Fortuitously, one of the prototypical ultra-active stars in the solar neighborhood,
AB~Doradus, at a distance of only d~=~15.3~pc from the Sun, is located at large ecliptic latitudes 
($\beta$ = - 86.6$^{\circ}$), and  for this reason AB~Dor receives rather extensive simultaneous exposure with the
eROSITA (see Sect.~\ref{sec_eros}) and TESS satellites (see Sect.~\ref{sec_tess}).  

The rotation period of  AB~Dor is very short (P = 0.51~d), and consequently it shows all the typical signatures of magnetic activity
such as star spots and chromospheric  and coronal emission, the signatures of which we explore in this
study.   AB~Dor~A is a member of a quadruple system
discussed in detail by \cite{guirado2011}, but the other system components are much fainter
and are ignored in the following.
AB~Dor was first detected as a strong X-ray source with the {\it Einstein} Observatory by \cite{pakull1981},
and since then has been observed with most subsequent X-ray satellites:  with ROSAT \citep{kuerster1997}, 
{\it XMM-Newton} (\cite{guedel2001}, \cite{lalitha2013a}), and {\it Chandra} \citep{hussain2007}.
In addition to showing strong ``persistent'' X-ray emission,
AB~Dor frequently flares at X-ray wavelengths, and a detailed
statistical study of these X-ray flares is presented by \cite{lalitha2016}, extending an earlier study
based on GINGA data by \cite{vilhu1993}.   
Most relevant for our study is that of \cite{kuerster1997},
who analyzed five years of ROSAT X-ray observations of AB~Dor, finding no evidence for long-term changes 
in its X-ray flux and only some partial modulation with its rotation.  \cite{lalitha2013b} analyze the
long-term X-ray behavior of AB~Dor from multi-satellite data and find that its persistent level of X-ray
emission does not show long-term changes by more than a factor of two, possibly modulated
along with its photometric variability.

Extensive ground-based photometry of AB~Dor taken 
over decades is available and has been presented by \cite{jaervinen2005}; one of the main results of this study
is the conclusion that the surface spots of AB~Dor are grouped around two active longitudes separated on average by $180^{\circ}$ and 
that they migrate with a variable rate because of surface 
differential rotation.  These studies show the persistent presence of large spots on the surface
of AB~Dor, which change in size and location over the years; for example, \cite{ioannidis2020}  used TESS data to study the spot surface 
distribution on AB~Dor in 2018 and also find (see their Fig.~4) two major
spots that are slowly changing their respective positions with time.

Photometric studies of active stars have been revolutionized by space-based
observatories such as CoRoT, {\it Kepler,} and TESS both in terms of achievable
photometric precision and temporal coverage, because they are not affected
by scintillation, day and night, weather, and other effects.
\cite{schmitt2019} present a detailed analysis
of optical flares recorded on AB~Dor during the first two months of TESS observations, and report
the occurrence of eight ``superflares'', that is, flares with a total energy release in excess
of 10$^{34}$ erg.  \cite{schmitt2019} show that the occurrence rate of these events is
about one per week, and further present a re-analysis of an X-ray flare on AB~Dor, which
was first described by \cite{lalitha2013a}), demonstrate the superflare nature of this event, and
show that the energy release at optical wavelengths likely exceeds that at
X-ray wavelengths.  \cite{ioannidis2020}, extending these studies by analyzing the
full set of TESS AB Dor data obtained during its first year of operations, 
analyzed ---among other things--- the spot distribution and its changes and the occurrence of flares
with respect to the spot distribution of AB~Dor.  In particular, the flare occurrence on AB~Dor appears to be
linked to the presence of the active regions on its visible surface, that is, the chance that a flare will occur
is larger when AB~Dor is photometrically fainter.

The purpose of this paper is to present the results of the first three eROSITA surveys on AB~Dor,
with particular emphasis on the long-term evolution of its  X-ray flux and the simultaneous
observations obtained between eROSITA  and TESS.  In section 2 we discuss basic properties
of the eROSITA and TESS satellites in as much they are relevant for our study, and in section
3 we present our new observational results on AB~Dor.  Section 4 contains a discussion and in section 5
we present our conclusions.

\section{eROSITA and TESS satellites and data analysis}
\label{sec_eros_tess}

\subsection{The eROSITA instrument on board SRG} 
\label{sec_eros}

The eROSITA instrument ({\bf e}xtended {\bf RO}entgen {\bf S}urvey with an 
{\bf I}maging {\bf T}elescope {\bf A}rray) is the soft X-ray  instrument on board
the Russian--German Spectrum-Roentgen-Gamma mission (SRG).  After its launch from Baikonur,
SRG was placed into a halo orbit around the Sun--Earth L2 point, where it is performing a four-year
all-sky survey, which started in December 2019.  The eROSITA all-sky 
survey is carried out in a way that is similar 
to the ROSAT all-sky survey, that is, the sky is scanned in great 
circles perpendicular to the plane of the ecliptic. 

The longitude of the scanned great circle, that is, the survey rate,  moves by $\sim$1$^{\circ}$ per day, 
and thus the whole sky is covered in half a year; this procedure will be
carried out eight times over the  life time of the mission, and we refer to the respective
data sets as eRASS1 to eRASS8. In this paper we present data from the surveys
eRASS1-3.

 The eROSITA  scan rate is set to 1.5~deg/min or one rotation in 4~hours. 
Thus, given the eROSITA field of view of $\sim$ 1 deg, a typical
single scan exposure can last up to 40~sec.   As a consequence of this survey geometry, the
periods of elapsed time during which a given source is observed depend sensitively
on its ecliptic latitude:  sources near the plane of the ecliptic are observed for
a day, and sources at high ecliptic latitudes such as AB~Dor are observed for a couple of weeks, but always with
a cadence of 4~hours.

The eROSITA instrument contains seven X-ray telescopes,
each equipped with their own CCD camera in the focal
plane of the respective telescope.  
All seven telescopes and cameras are operated independently, but they all look parallel.  
Except for some slight differences in filters, the seven units are identical, thus providing 
a high degree of redundancy.
The  energy range eROSITA is between 0.2~and ~8~keV, 
its spectral resolution approaches 80~eV at an energy of 1~keV, 
which is significantly higher than that obtained by the ROSAT PSPC and 
opens up entirely new scientific investigations; a detailed description of the
eROSITA hardware, mission, and in-orbit performance is presented by \cite{predehl2021}.

\subsection{The TESS satellite} 
\label{sec_tess}

The Transiting Exoplanet Survey Satellite (TESS) was launched on April 18, 2018, 
into a highly elliptical 13.7~day orbit in a 2:1 resonance with the Moon.  This type of orbit
provides a very stable environment for its operations, but it also leads to interruptions
in the data stream with the same period;  a detailed description of the 
TESS mission is given by \cite{ricker2015}. The primary scientific goal of the
TESS mission is the detection of exoplanets around the brighter stars using the transit method, 
and to this end TESS performs differential photometry for many thousands of stars.
TESS is equipped with four red-band optimized wide-field cameras that continuously 
observe a 24$^{\circ}$ $\times$  96$^{\circ}$ strip on the sky for 27 days.  Subsequently,
the TESS pointing is moved by about 24$^{\circ}$  along the plane of the
ecliptic, and thus a whole hemisphere is covered within a year.  In subsequent years observations
switched between the northern and southern ecliptic halves of the sky.
TESS began its observations on the southern hemisphere prior to the launch of
eROSITA.  When the eROSITA all-sky survey began in December 2019, TESS was still observing
north, but switched then to observations of the southern hemisphere.

\subsection{Data analysis: eROSITA} 

For our eROSITA  study of AB~Dor, we use the results of the first three of eROSITA's eight planned all-sky surveys, 
using the processing results of the eSASS pipeline (currently version 946); a detailed description
of this software and the algorithms used is given by \cite{brunner2021}. Briefly,
the eSASS system analyzes the incoming photon stream, attaches a sky position and energy to each recorded 
photon, and sorts the events into sky tiles with a size of 3.6$^{\circ}$ $\times$ 3.6$^{\circ}$.

The events organized in sky tiles can be further analyzed, and  with the program {\it scrtool} photons around a given position
can be extracted and light curves constructed including all necessary corrections such as vignetting and dead time.  For the  
studies reported in this paper we used the nominal source of position of AB~Dor and
an extraction circle with a radius of 72 arcsec to gather all the source photons.  
The effective area of eROSITA drops significantly for energies in excess of about 2.3~keV, and therefore we only
consider events in the energy range 0.2~keV~-~2.3~keV.   To convert from count rate to energy flux,  we need
an energy conversion factor of (ECF), which depends on the spectral composition of the incident X-ray flux;
the derivation of the ECF as used in this paper is presented in Sect.~\ref{erosecf}.
We note in this context that all count rates reported in this paper refer to equivalent on-axis count rates for all seven telescopes:
due to the scanning nature of eROSITA, even the count rate of a constant source is changing all the time, and the
concept of a ``mean'' observed count rate is not useful, because it would depend on the exposure history of the source,
the number of telescopes used, and so on.

\subsection{Data analysis: TESS} 

The TESS project provides the light curves of pre-selected
targets (which AB~Dor belongs to) with a cadence of 2~minutes;
these light curves were generated with the so-called
Science Processing Operations Center (SPOC) pipeline,
which was derived from the corresponding pipeline for the
{\it Kepler} mission \citep{jenkins2016};  further information
can be found at the website at \url{https://heasarc.gsfc.nasa.gov/docs/tess/pipeline.html}.
The light curves are publicly accessible and available for
download from the Mikulski Archive for Space Telescopes (MAST)\footnote{https://mast.stsci.edu}.
We specifically use the so-called Simple Aperture Photometry (SAP) fluxes for our analysis.

\begin{figure}  
\begin{minipage}{0.45\textwidth} 
\includegraphics[width=\linewidth]{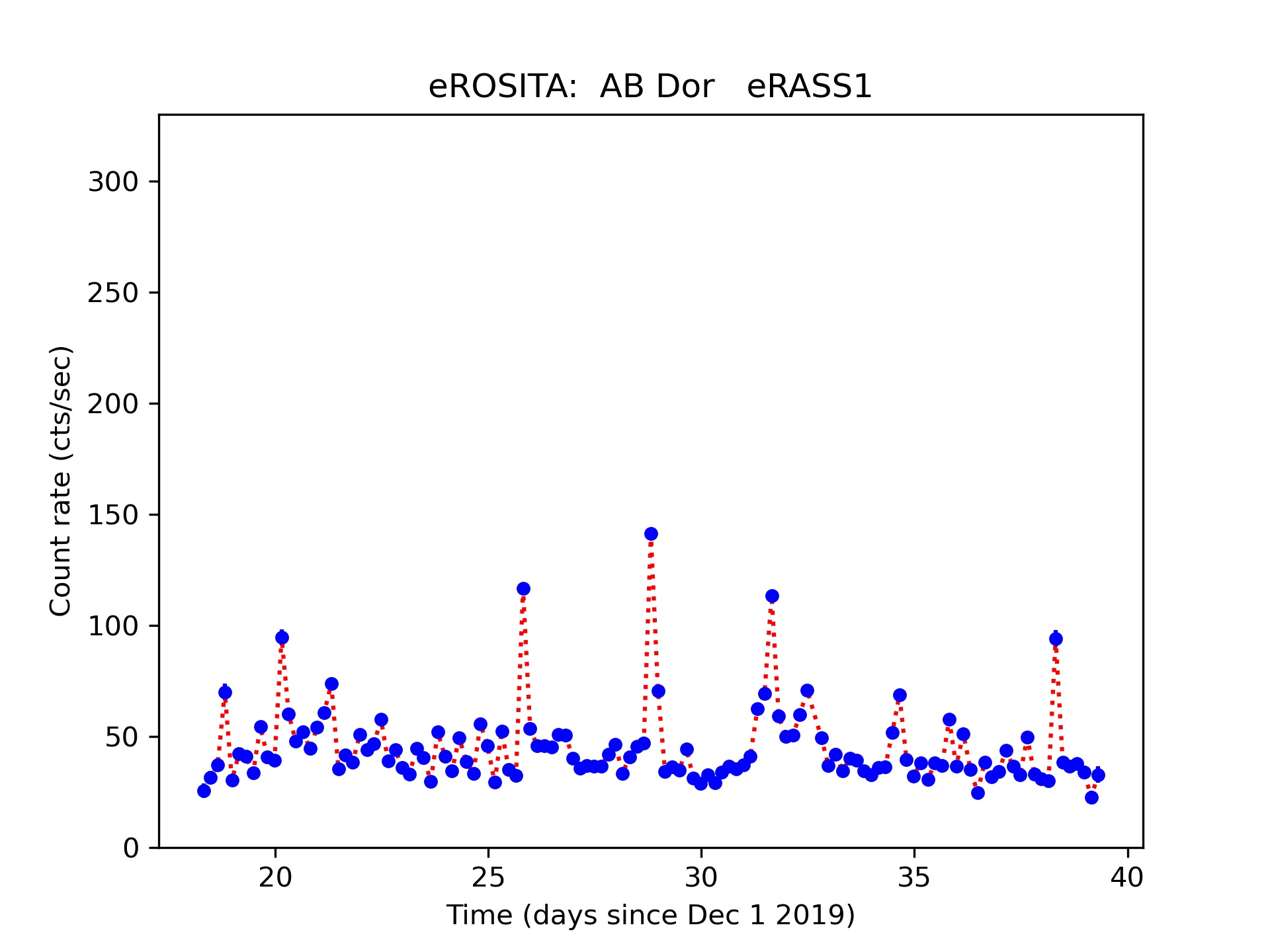}   
\caption{Scan-averaged AB Dor light curve obtained during eRASS1 in December 2019 and January 2020; blue dots: eROSITA data points,
dashed red line connects adjacent data points for better visibility.\label{abdor_lc1}} 
\end{minipage}    
%\hspace{\fill}  %% no blank line before of after this instruction
\vspace{0.75cm}
\begin{minipage}{0.45\textwidth} 
\includegraphics[width=\linewidth]{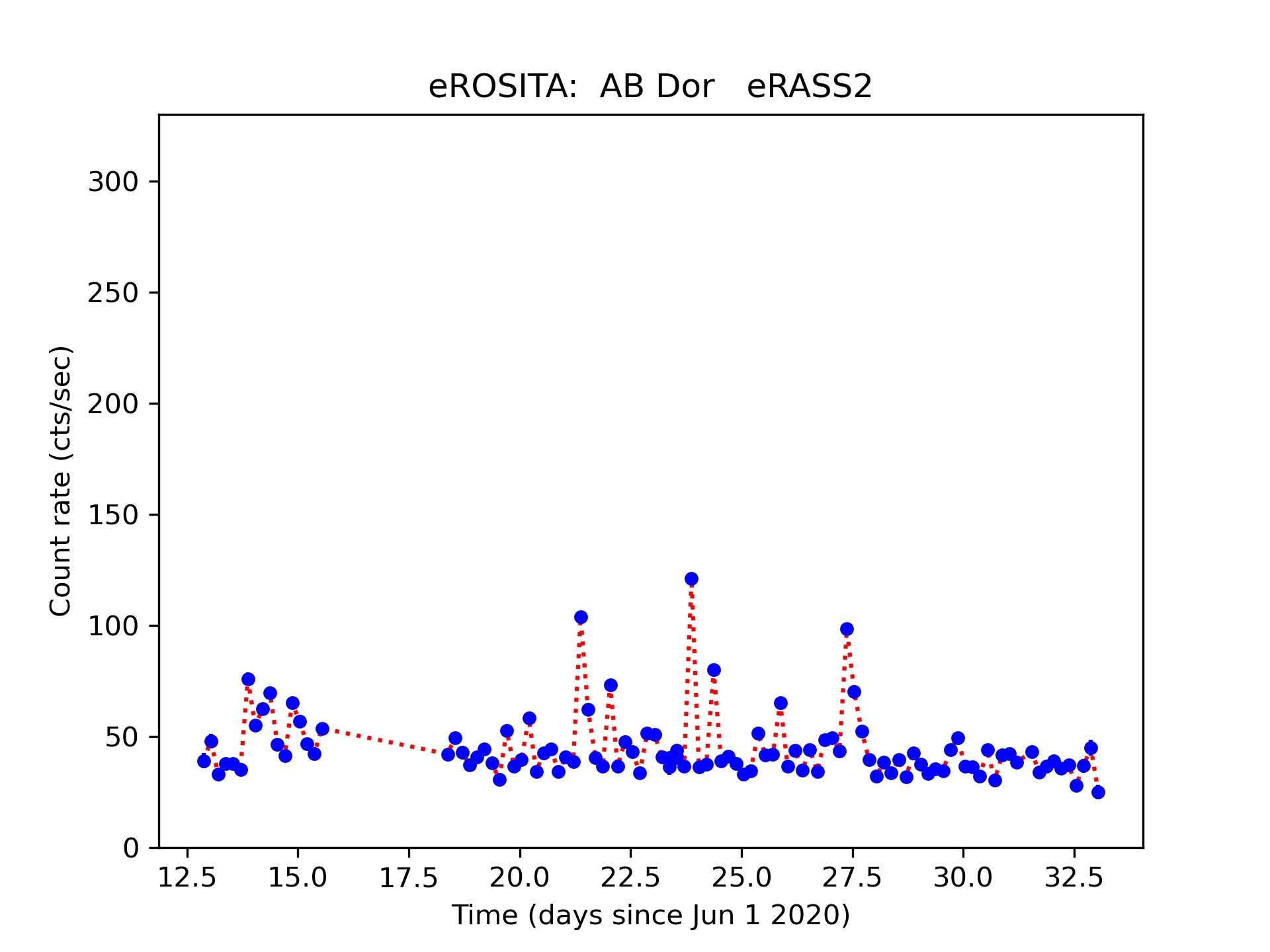}    
 \caption{Scan-averaged AB Dor light curve obtained during eRASS2 in June and July 2020: blue dots: eROSITA data points,
dashed red line connects adjacent data points for better visibility.\label{abdor_lc2}}
\end{minipage}    
\begin{minipage}{0.45\textwidth} 
\includegraphics[width=\linewidth]{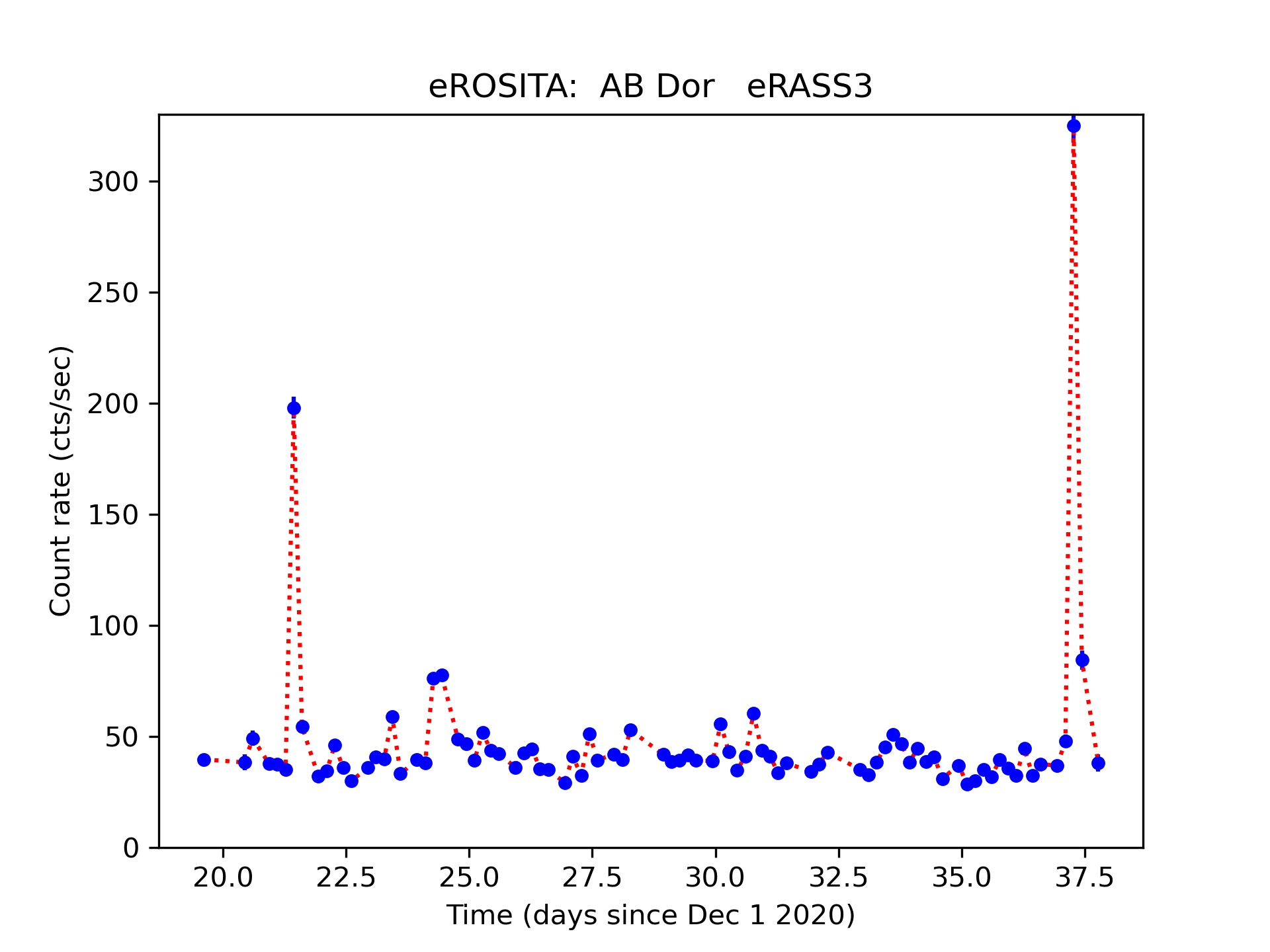}   
\caption{Scan-averaged AB Dor light curve obtained during eRASS3 in December 2020 and January 2021; blue dots: eROSITA data points,
dashed red line connects adjacent data points for better visibility.\label{abdor_lc3}}
\end{minipage}   

\end{figure}

\section{Results}

\subsection{eROSITA light curves for AB~Dor} 

In Figs.~\ref{abdor_lc1} -~\ref{abdor_lc3} we show the eROSITA light curves measured for AB~Dor
during the eRASS1, eRASS2, and eRASS3 observations; we note that the formal count rate errors 
are typically smaller than the plot symbols in these figures.  Being a strong source with typically (at least) 
$\sim$ 40~cts/sec, AB~Dor can be detected in almost all scans. To obtain results with
reasonable statistics, we only consider scans with at least five counts and more than 3 seconds of
vignetting-corrected exposure, and end up with a total of 319 valid scans obtained during
the eRASS1-eRASS3 observations.   As discussed above, eROSITA provides ``snapshots'' of AB~Dor's corona
every 4~hours during a period of almost three weeks.   Given AB~Dor's rotation period of  slightly over 12~hours, the star
rotates by 120$^{\circ}$ between two subsequent scans, and after three scans eROSITA views (almost) the
same stellar hemisphere again.   This implies that subsequent scans represent almost independent
samples of the coronal state of AB~Dor -- at least for coronal regions near the surface.
 
\begin{center}
\begin{figure}[h]
\includegraphics[width=9.0cm]{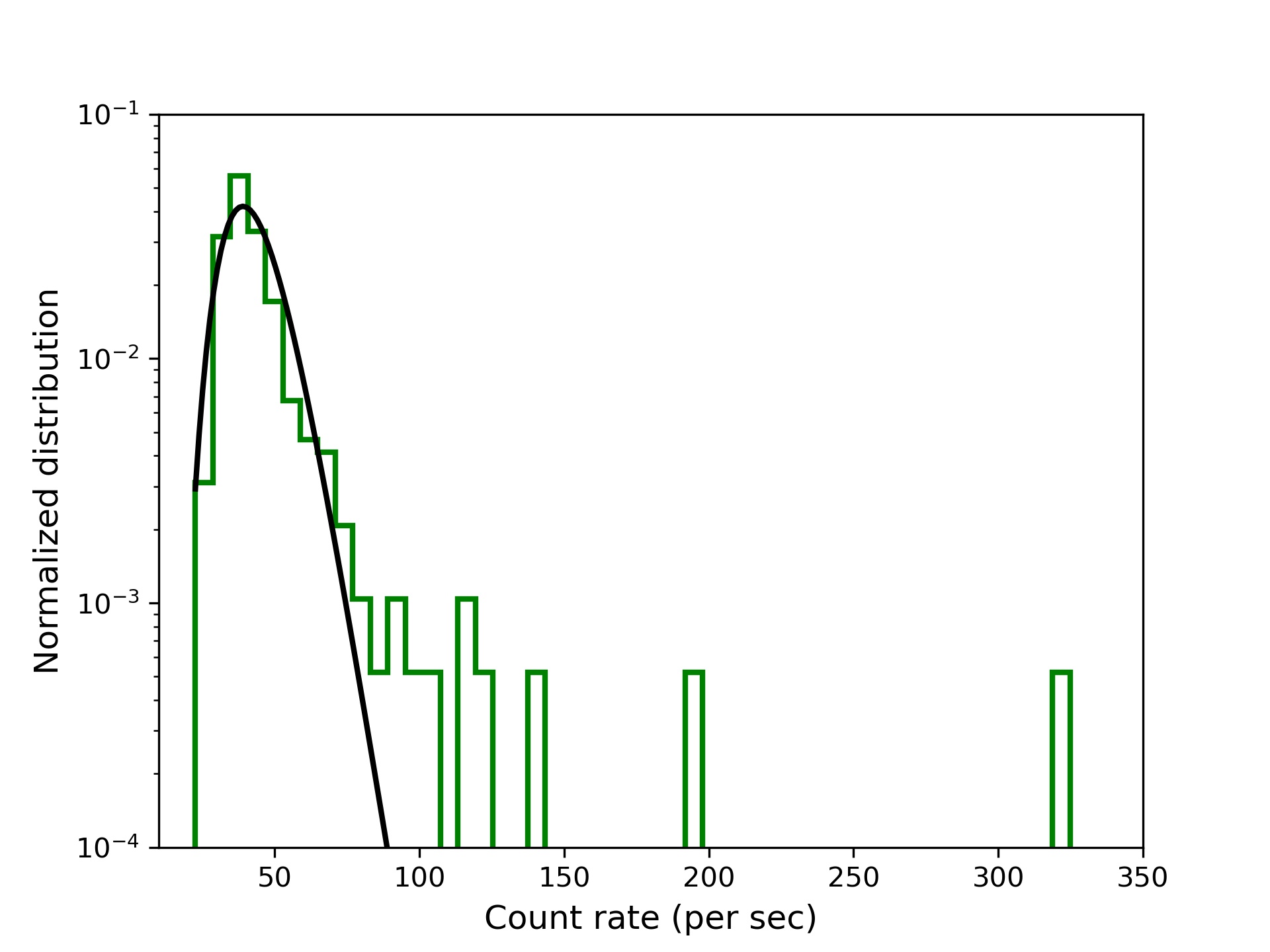}
\caption{Histogram of AB Dor scan-averaged count rates obtained during eRASS1-3. \label{ratehist}}
\end{figure}
\end{center}

Variability is a clear characteristic of AB Dor's X-ray light curve:  
arbitrarily defining every scan with count rates in excess of 90~cts/sec as a ``flare'', 10 out of the observed
319 scans fall into this category.
On the other hand, once flares are excluded, AB~Dor appears to have some ``memory'' as to what its count rate
and X-ray flux  should be.  To quantify the encountered count rate distribution, in Fig.~\ref{ratehist} we show 
a histogram of all valid AB~Dor eROSITA scans recorded in the surveys eRASS1-3, together with a log-normal 
distribution fit to the bulk part of the rate distribution below 100 cts/s;  the best fit parameters are ln($\mu_{rate}$) = 3.72 and  $\sigma$ = 0.23.
Around 5~\% of the distribution are outside the log-normal distribution, and the largest events are discussed in
Sect.~\ref{sec_flares}.

\begin{center}
\begin{figure}[h]
\includegraphics[width=7.0cm]{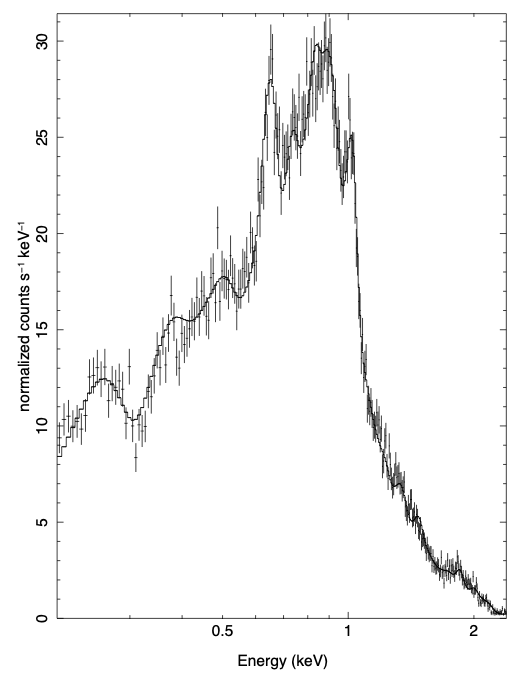}
\caption{Scan-averaged quiescent X-ray spectrum of AB~Dor (black data points) together with a model fit consisting of two isothermal plasma emission
models (solid line); see text for more details.
\label{erosspec}}
\end{figure}
\end{center}

\subsection{eROSITA spectrum for AB~Dor} 
\label{erosecf}

To illustrate the spectral properties of AB~Dor as measured by eROSITA, we present in Fig.~\ref{erosspec} a spectrum of the quiescent emission
of AB~Dor together with a spectral fit composed of two optically thin temperature components with temperatures of (0.66$\pm$0.01)~keV and (2.03$\pm$0.14)~keV
and variable abundances, which was obtained using the {\it XSPEC} package 
\citep{arnaud1996}.   Figure~\ref{erosspec}  shows that the X-ray flux of AB~Dor as measured by eROSITA is concentrated in the spectral range
up to 2.3 keV; the rapid drop in X-ray flux seen between 1~keV and 2~keV is both due to the intrinsic properties of AB Dor's corona and the rapidly
falling effective area of eROSITA.  One also notices a strong line near 0.67~keV, which is caused by OVIII~Ly$\alpha,$ as well as a much
weaker ``line'' near 0.525~keV, the so-called OVII triplet; one further notices the Ne~IX triplet near 1.08 ~keV and the Fe~L complex
between 0.8 and 0.9~keV.   The plasma model fits the observed spectrum quite well; the formal
value of $\chi^2$ is 1.19. The ratio between the model flux (in the spectral range 0.2 - 2.3~keV) and the vignetting-corrected count rate
is 1 $\times$ 10$^{-12}$ erg/cm$^2$/ct, which is the required ECF to convert from count rate to energy flux.  It is difficult to estimate the
accuracy of this ECF. A major source of uncertainty is the vignetting correction which depends both on energy and the off-axis angle at
which a photon is registered, yet these uncertainties ought to be below 10\%.  This ECF was used to convert from measured count rate to 
inferred energy flux for all the AB~Dor scans obtained by eROSITA.

\subsection{TESS light curves for AB~Dor} 
\label{tess_lc}

\begin{center}
\begin{figure}[h]
\includegraphics[width=9.0cm]{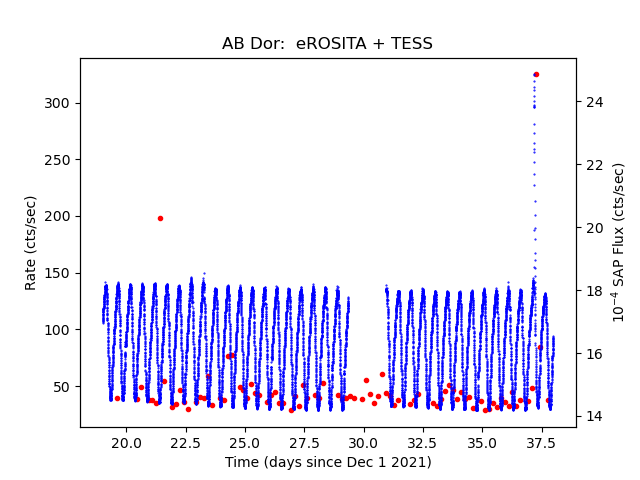}
\caption{TESS light curve for AB Dor (blue data points, right ordinate) concurrent with eROSITA's eRASS3 survey observations (red data points, left ordinate). 
\label{tesslc}}
\end{figure}
\end{center}

In the following section we discuss the light curves obtained by TESS concurrent with 
eROSITA's coverage of AB~Dor during the eRASS3 observations.  In 
 Fig.~\ref{tesslc} (labels on right y-axis) we plot the TESS SAP flux as a function of time (blue data
 points). The comparatively much more sparsely sampled eROSITA rates are also plotted for comparison (red data points).
 The TESS light curve is characterized by its substantial modulation interpreted as rotational
 modulation through star spots. From late December 2020 to early January 2021, the typical peak-to-peak variation with respect to the mean flux is found to
 be 22\%, quite a bit larger than the 18\% peak-to-peak variations found by \cite{schmitt2019}. 
 
 A very noticeable feature is the huge flare that occurred on Jan 7$^{}$, 2021, peaking at around UT~4:15 and analyzed below.
 We also phased the TESS light curve of AB~Dor folded with a period of 0.514275~days (i.e.,
 the rotation period of AB~Dor derived by \cite{schmitt2019} from TESS data obtained in 2018) and found
 that this value also provides a good description of the light curve data shown in
 Fig.~\ref{tesslc}.   Furthermore, our phasing analysis (cf., Fig.~\ref{phasedlc}) also shows that there is very little
 light curve evolution, which turns out to be extremely useful for the analysis of the large flare.
 
 \begin{center}
\begin{figure}[h]
\includegraphics[width=9.0cm]{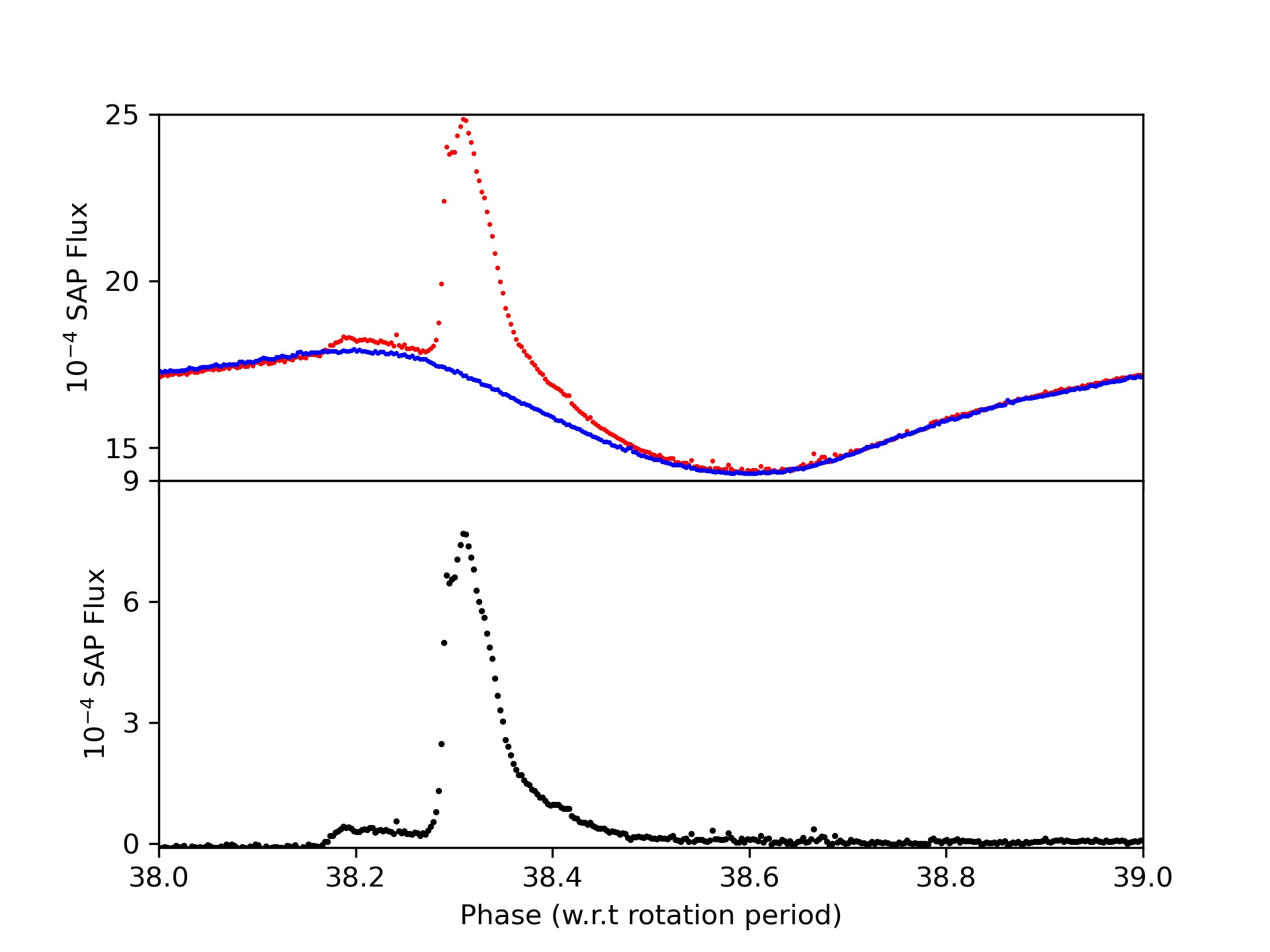}
\caption{TESS light curve (in terms of 10$^{-4}$ SAP flux) of the large flare on AB Dor, phased 
with rotation period. Upper panel:  Blue data points  are the mean of the light curves in rotations before and after the flare, and the    
red data points  are the rotation during the flare. Lower panel: Net flare light curve. \label{tessflare}}
\end{figure}
\end{center}

 In Fig.~\ref{tessflare} we show some phased TESS light curves obtained around the big flare; we note that the phase interval of unity 
 corresponds to 0.514275~days. In the upper panel of Fig.~\ref{tessflare} the light curve during the flare (red data points)
 is shown as well as the light curve averaged over the rotation prior to and after the flare; fortuitously
 the light curves  before and after the flare are extremely similar, which leads us to assume
 that the (background) light curve during the flare has also not changed.  Modeling this
 background as the mean of the light curves before and after, we can subtract the nonflare background and
 obtain the pure flare light curve, also shown in Fig.~\ref{tessflare} (black data points in lower panel).
 As demonstrated by Fig.~\ref{tessflare}, the whole flare event lasts a little over seven hours.
 Prior to the actual impulsive event, there appears to be some pre-flare activity lasting almost three
 hours.  The actual rise to flare peak then occurs in about 30 minutes. There appears to
 be some substructure in the flare rise phase, consisting possibly of two events with shorter rise times.
 After the flare peak, the light curve decreases more or less exponentially and merges
 into the photospheric background after about 3.5 hours, suggesting that the whole flare duration is about
 seven hours.
 
 This overall flare duration of 7-8 hours is considerably longer than AB~Dor's half rotation period of 6.1 hours.   
 Doppler imaging (\cite{kuerster1994}, \cite{kuerster1996})
suggests an inclination angle of AB~Dor of around 60$^{\circ}$, and therefore on the stellar surface there is a continuity 
from regions never visible around the
``south pole'' to regions always visible around the ``north'' pole.  As there are no  features in the light curve that
suggest rotational modulation of the flare emission, the flare site was likely at some higher latitude,
but photometry alone is insufficient to locate the flare site.
 
 \subsection{Analysis of TESS flares}
\label{tess_resp}

\subsubsection{Energetics}

To determine the energetics of the flare, we follow the approach
outlined by \cite{schmitt2019}.   As discussed by  \cite{schmitt2019},
the spectral band pass of TESS is relatively broad (it comprises the
wavelength range between 6000~\AA \,and 10000~\AA \,centered on the I-band)
and captures only some part of the overall flux.    For simplicity, we assume
blackbody spectra, although the photospheric spectrum of AB~Dor
is clearly  not that of a blackbody, and once the flare sets in, the temperature
is continuously varying, yet the TESS data do not allow
us to diagnose these changes.
To convert the instrumental TESS fluxes into physical values, we assume a bolometric
luminosity for AB~Dor of 1.5~$\times$10$^{33}$~erg/sec, an effective temperature of
5100~K (cf., \cite{close2007}) and a distance of 15.3~pc;  in 
the literature one finds a range of possible effective temperatures, yet those differences 
lead to rather small changes in the conversion from SAP rates to energy fluxes (in the TESS band).

However, the TESS band captures only part of AB~Dor's overall emission.
The fraction of the total flux captured by TESS depends on the temperature and reaches values
of $\approx$ 0.35 for temperatures near 5000~K.
For hotter plasma temperatures, as expected for the flaring plasma, more and
more flux is emitted outside the TESS band, and therefore that fraction decreases. By
numerically integrating the recorded flare light curve
(shown in Fig.~\ref{tessflare}), we can determine the total SAP count of the flare.
To accurately convert SAP count into fluence, we would have to know
the evolution of flare temperature with time which is unknown.   To stay on the
conservative side, we therefore
apply the derived photospheric energy conversion factor  and
estimate a total optical flare energy of 4 ~$\times$~10$^{36}$ erg, which is an order of 
magnitude higher than the most powerful AB~Dor flare discussed by \cite{schmitt2019}.  
We stress that this number is ---in the context of our approximations--- a lower limit
to the actual energy release because the flare temperature always exceeds the
photospheric temperature and the energy radiated in the rise phase is missing entirely
from our considerations.

\begin{center}
\begin{figure}[h]
\includegraphics[width=9.0cm]{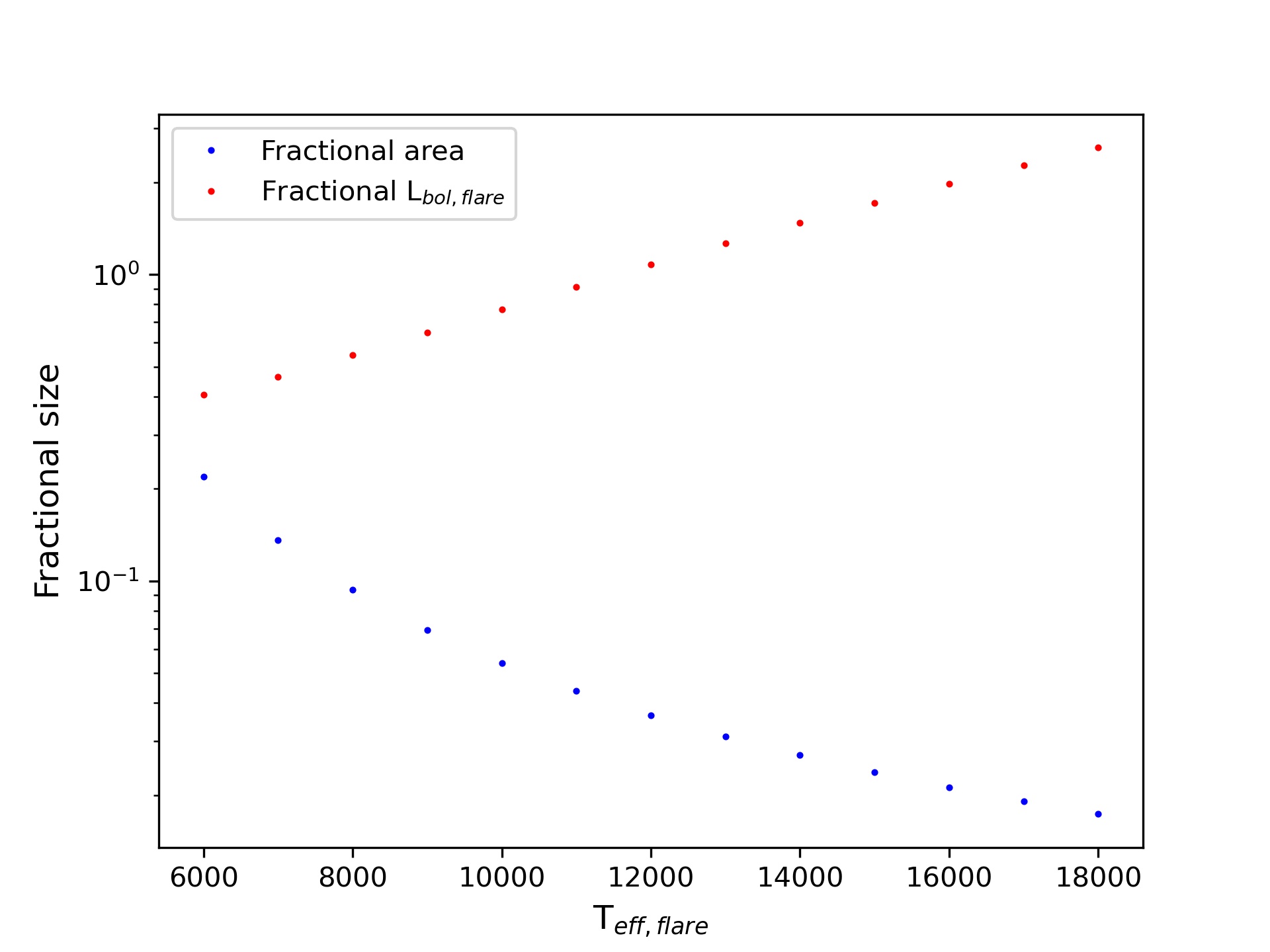}
\caption{Computed flare filling factor scaled by the surface area of  AB~Dor (blue data points) and
flare bolometric luminosity (at observed flare peak)  scaled by its bolometric
luminosity (red data points); see text for more details.
\label{fillfac}}
\end{figure}
\end{center}

\subsubsection{Flare filling factor}
\label{sec_fill}

Again following  the analysis of \cite{schmitt2019}, we can estimate 
the filling factor of the flare by computing the emitting area from Stefan-Boltzmann's law,
the determined peak luminosity and an assumed flare temperature.   At optical flare peak,
the plasma temperature was (presumably) much higher, it is therefore interesting to
compute these values as a function of the unknown flare effective temperature.
In Fig.~\ref{fillfac}, we present the bolometric instantaneous luminosity of the flare (scaled
by the quiescent bolometric luminosity of AB~Dor) and compute the flare filling
factor (scaled by AB~Dor's surface) as a function of effective temperature.
Figure~\ref{fillfac} shows that for flare temperatures above about 12000 K, the
bolometric flare luminosity is equal to AB~Dor's bolometric luminosity, while at the same
time the flare filling factor is only a few percent.   This is a consequence of the
T$^4$ dependence of Stefan-Boltzmann's law.  Needless to say, these numbers should be considered
as magnitude estimates in view of all the existing uncertainties, yet the
calculated flaring area is still small compared to the stellar radius and the overall flare area
is still in the one-digit percentage range.   In contrast,  the spot filling factors  required to explain 
the observed rotational modulations are quite large based on the empirical fitting 
formulae provided by \cite{maehara2017} as well as the explicit spot modeling of AB~Dor
performed by \cite{ioannidis2020}.   Using the formulae provided by \cite{maehara2017}, we find
a spot filling factor of between 10\% and 20\% for the whole star, and \cite{ioannidis2020} require three
large spots with typical sizes of 0.15-0.2 R$_{*}$ (cf., Fig. 4 and 5 \cite{ioannidis2020} )
in terms of stellar radius to account for the measured brightness variations;  we note that for all 
photospheric modeling, we must assume a temperature difference between a spot and the pristine
photosphere.
 
\begin{center}
\begin{figure}[h]
\includegraphics[width=9.0cm]{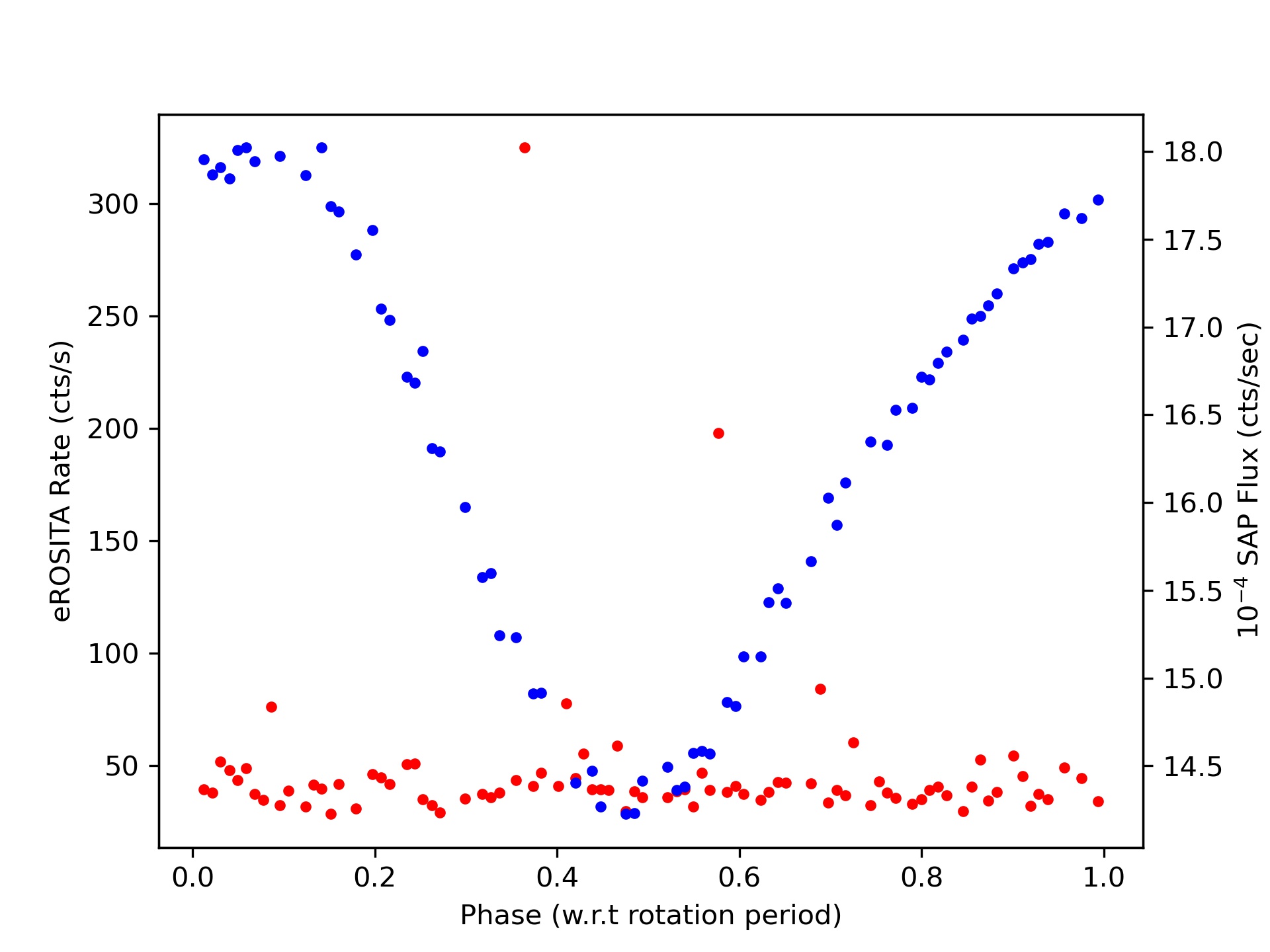}
\caption{Phased light curve of concurrent eROSITA and TESS data for AB Dor; red data points show the eROSITA
count rates (left hand axis), blue data points show the TESS SAP fluxes (right hand axis). \label{phasedlc}}
\end{figure}
\end{center}

\subsection{Joint analysis of eROSITA and TESS data} 

We now focus on the data simultaneously taken by the eROSITA and TESS missions, and discuss the properties of
AB~Dor's quiescent and flaring emission.

\subsubsection{Quiescent emission}

In Fig.~\ref{phasedlc} we plot the eROSITA (red data points) and TESS light curves (blue data points) obtained during
the concurrent eRASS3 observations. We note that the TESS flare data points (shown in Fig.~\ref{tessflare}) have been left out,
while all eROSITA data are shown. To produce the TESS data points, we interpolated between the available measurements
around the time for each eROSITA scan, thus producing a TESS light curve with a spacing of four hours. We
also note that for better visibility a constant offset was removed from the TESS data and the subtracted light curve was
scaled by a factor of 100.   One first notes very little optical light curve evolution during the three weeks of
concurrent data taking, that is, during that period the spot configuration on AB~Dor stayed more or less constant.
Two further X-ray flares can be seen, one near phase 0.62, having occurred on Dec~22 at 10:32, and 
another one near phase 0.42, having occurred on  Jan~7 at 6:32.; these flares are discussed in Sect.~\ref{sec_flares}.  
In contrast to the optical light curve, the
X-ray light curve shows no apparent structure and a correlation analysis invoking Pearson's r as well as
a Spearman's rank correlation coefficient formally confirms that optical and X-ray data are uncorrelated, implying that
the X-ray emission ---at least as observed during eRASS1-3--- does not depend on the rotational phase and therefore
does not depend on the spot coverage over the visible hemisphere of AB~Dor.   As a consequence, it is quite difficult
to directly determine rotational modulation from X-ray data for active stars, while such determinations are (usually) straightforward from
optical monitoring data.

\begin{center}
\begin{figure}[h]
\includegraphics[width=9.0cm]{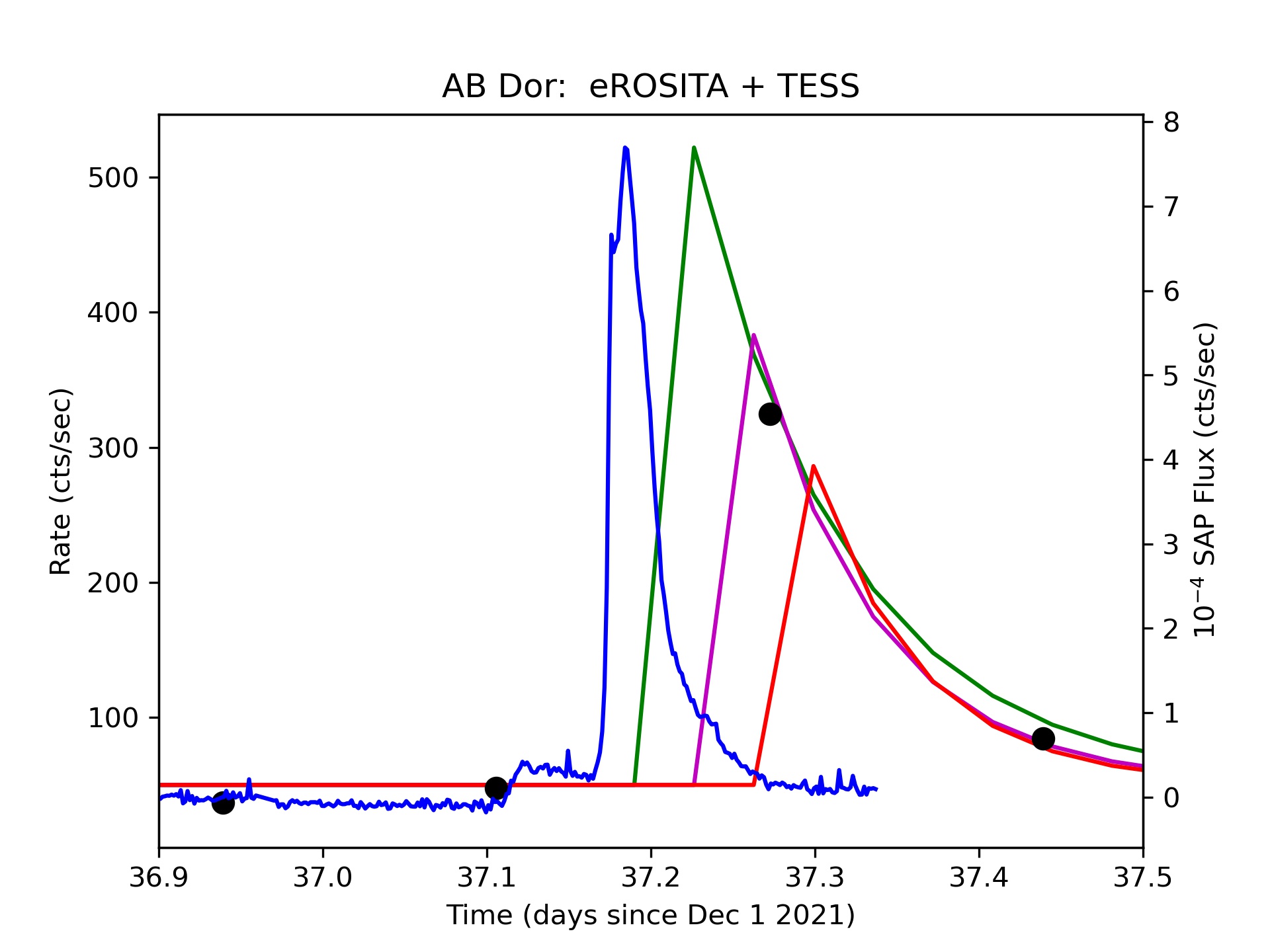}
\caption{eROSITA X-ray observations (black data points) and TESS flare light 
curve (blue solid line).  Some hypothetical X-ray flare light curves are shown in the colors red, magenta, and green.
Red corresponds to a minimal-energy scenario and green is close to a most likely scenario; see text for 
more details. The left and right axes apply to the X-ray and TESS data, respectively.
\label{flaremodel}}
\end{figure}
\end{center}

%\subsection{TESS light curves for AB~Dor} 

\subsubsection{Flaring emission}
\label{sec_flares}

In the joint eROSITA/TESS light curve of AB~Dor (shown in Fig.~\ref{abdor_lc3} and Fig.~\ref{tesslc})
there are two occasions when eROSITA shows count rates in excess of 100~cts/s: one on Dec~22 at 10:32  and 
another on Jan~7 at 6:32.  The latter
event, which is so far the  event with the largest count rate observed by eROSITA, is clearly accompanied by an
equally huge TESS flare of almost twice the size of the rotational modulation.

For the first of these flares, which is very noticeable in X-rays, there is no conspicuous flare event in the TESS
light curve.  A very small flare with a duration of less than 8 minutes is observed about 40 minutes prior to the time of the
eROSITA X-ray scan.  Whether or not these events are related to each other cannot be deduced from the
available data.  However, we can say with confidence that the X-ray flare is not accompanied by any
conspicuous energetic optical event.

 The situation is very much different for the second flare on Jan 7$^{}$. As evident from Fig.~\ref{tessflare}, the
X-ray flare does occur during a major optical flare, and given that the scan  4 hours after the peak is still 
elevated, it is tempting to interpret the X-ray light curve as arising
from a long-duration flare event.  

The four-hour temporal cadence of eROSITA survey observations 
is not that well suited for the study of stellar flares.  While ideally the X-ray light curve should be continuously covered,
for an assessment of the X-ray energetics one needs the peak amplitude A$_{\mathrm{peak}}$ and the flare
decay time $\tau_{\mathrm{decay}}$, because for an assumed exponential decay the total X-ray energy is proportional
to A$_{\mathrm{peak}} \times \tau_{\mathrm{decay}}$.   Furthermore, it is helpful to know the start  times of flares and the times of their peaks
in order to obtain an assessment of the rise phase, which, however, normally contributes relatively little to the
overall energy budget because the rise times tend to be very short.  Unfortunately, none of these parameters
are accurately known, and therefore a precise determination of the X-ray energetics is impossible.

Given the extreme nature of the optical flare, the question arises as to what a reasonable X-ray scenario
might look like.   Given that the optical decay is quite exponential without any substructure, we can assume
that the X-ray decay is also purely exponential; as the X-ray rise phase is unobserved, we ignore it in our
``modeling''.   Needless to say, there are no unique solutions, and to give an idea of what the flare light curve
could have looked like, we present a few examples in Fig.~\ref{flaremodel}, where we show the measured
eROSITA data (blue points) and the derived TESS flare light curve (scaled by a factor of ten, blue curve).
We also show three hypothetical cases, that is, an X-ray flare peak close to the optical flare peak (green curve),
an X-ray flare peak close to the observed scan maximum (red curve), and a case in between (magenta).
For the simplified chosen models we computed total X-ray outputs of 1.2 $\times$ 10$^{35}$ erg,
6.4 $\times$ 10$^{34}$ erg, and 4.9 $\times$ 10$^{34}$ erg for the green, magenta, and red model curves.
Experience shows that the rise phase could contribute an additional 10\% or so to the total X-ray output,
yet it is difficult to see how any realistic flare scenario would yield an X-ray energy output significantly above
2 $\times$ 10$^{35}$ erg.

\begin{center}
\begin{figure}[h]
\includegraphics[width=9.0cm]{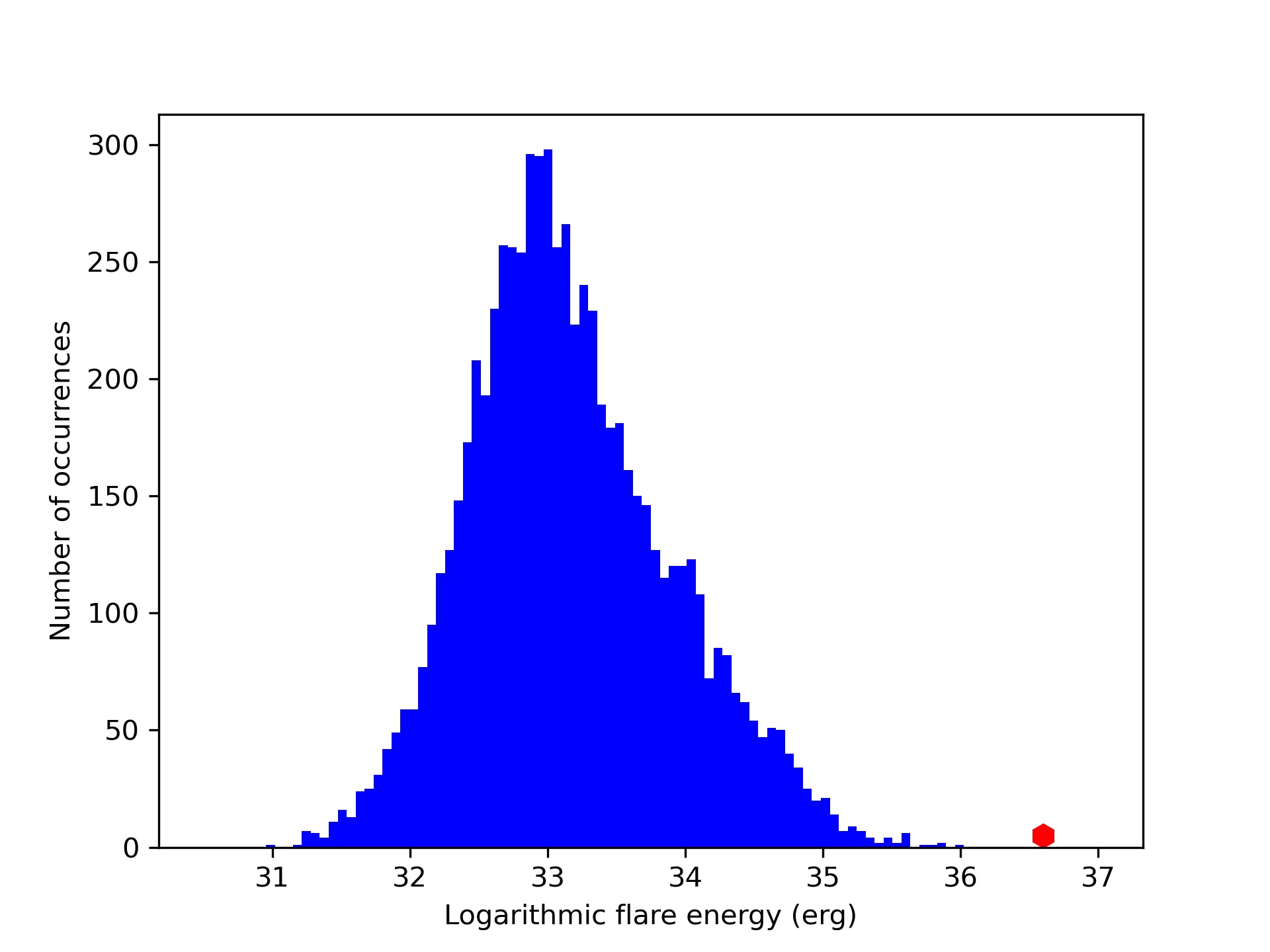}
\caption{Histogram of total flare energy outputs from TESS measurements of 7074 flares on late-type stars
(blue filled histogram, data taken from  \cite{guenther2020}) and the AB~Dor flare from 
Jan 7$^{}$ (red hexagon, this paper); see text for more details.
\label{flarehist}}
\end{figure}
\end{center}

\section{Discussion}
\label{sec_discussion}

\subsection{Quiescent X-ray and optical emission}

A general characteristic of AB~Dor's optical emission is the large modulations observed at almost all times as evidenced by the TESS light curves presented in this paper and those
by \cite{schmitt2019} and \cite{ioannidis2020}.   This finding demonstrates the presence of
large but inhomogeneously distributed star spots on AB~Dor's surface at all times.   The  X-ray flux from AB~Dor
as observed by eROSITA is also variable, however the kind of observed variability is different.
Once obvious flare scans are removed, the X-ray snapshots taken by eROSITA can
be well characterized by a log-normal distribution which appears to be constant at least
over the 1.5 years of eROSITA observations.  In contrast to the optical flux, no
significant modulation of the X-ray flux with the rotation period is detected, which one would naively expect
by assuming a correlation between larger spot coverage and stronger X-ray emission.
The coronae of rapid rotators such as AB~Dor should be saturated (for a discussion, see \cite{guedel2004}),
and indeed, computing the logarithmic ratio of X-ray and bolometric luminosities for AB~Dor, we find
log (L$_X$/L$_{bol}$) = - 3.17. Therefore, AB~Dor's corona is saturated as expected, and we suspect that
this may be responsible for the lack of observed X-ray modulations.

\subsection{eROSITA and TESS flares}

Both the TESS optical light curves (cf., \cite{schmitt2019}, \cite{ioannidis2020})
and eROSITA X-ray light curves show many flares (cf., Fig.~\ref{abdor_lc1} - \ref{abdor_lc3}), which 
we discuss in this section.
  
 \subsubsection{Comparison to optical flares on late-type stars}

For the flare on AB~Dor on Jan~7$^{}$ we determine a (minimal) optical energy of 4  $\times$ 10$^{36}$ erg;
to the best of our knowledge this is the most energetic flare event so far observed on AB~Dor.  
Energy releases on that order of magnitude are quite rare, to say the least.  
 \cite{guenther2020} present the results of a specific survey to detect
flares on late-type stars using the first months of TESS data, and report the detection of 
8695 flares in total, which occurred on 1228 individual
stars, out of which 673 are M dwarfs; the relevant data are listed in Table~1 and Table~2 in
 \cite{guenther2020}.  Out of the 8695 flares reported by  \cite{guenther2020},
only 11 had total energy releases of above 10$^{36}$ ergs.  Closer inspection of the
host stars for these large flares reveals a somewhat mixed picture, yet none of these stars
appear to be on or near the main sequence; also, 4 of the 11 flares above 10$^{36}$~erg
occurred on the star TIC332487879 (= HD~220096), which again is located far away from the
main sequence.  If we then restrict attention to late-type stars on or near the main sequence by
arbitrarily demanding stellar radii below 1.5 R$_{\sun}$, we end up with 7074 flares with
known energy releases as determined by TESS.   The histogram of these flare energy releases
 is shown in Fig.~\ref{flarehist}, together with the position of the large flare on AB~Dor.   

Figure~\ref{flarehist}  clearly demonstrates the unique
character of the event from Jan 7$^{}$ from the point of view of energetics; we note that all of
the eight superflares on AB~Dor
described by \cite{schmitt2019} are located in the bulk of the 
distribution shown in Fig.~\ref{flarehist} , emphasizing our point
that the flare from Jan 7$^{}$ represents a very unique event not only for AB~Dor.
This statement does not imply that K/M dwarfs cannot produce flares with
energy releases above 10$^{36}$ ergs; for example, using ground-based observations 
with the NGST, \cite{jackman2019}, report the detection of
a flare with an energy release of $\approx$ 3 $\times$ 10$^{36}$ erg on a pre-main sequence star
at a distance of 210~pc, and \cite{davenport2016} report flares observed by the {\it Kepler} telescope
in a sample biased towards G-type stars, some of which produce flare energy releases up to 10$^{37}$ erg,
albeit on rather distant stars.  This is in line with the findings of \cite{wu2015}, also based on
{\it Kepler} studies of G-type stars, who report flare energy releases in the range  10$^{36}$ erg - 10$^{37}$ erg
in their sample.
 
 \subsubsection{Flares simultaneously observed by eROSITA and TESS} 
 
 Focusing now on the data simultaneously covered by eROSITA and TESS, there are two X-ray flares 
 with scan-averaged count rates in excess of 100 cts/s or L$_X > $ 2.5 $\times$ 10$^{30}$ erg/s.  
While the X-ray flare on Dec~22$^{}$ at 10:32 shows very little or no optical emission, the opposite holds for the
Jan~7$^{}$  flare at 6:32.  We visually inspected the TESS light curve and can state with certainty that there are no
 additional major flares to the Jan 7  flare$^{}$ described in Sect.~\ref{sec_flares}. 

Clearly, the temporal sampling of the X-ray light curve is somewhat sparse, and so we are unsure of how certain we can be about the derived 
energetics.  In this context, we again note that the total energy
release of a flare is essentially determined by only two numbers, the peak amplitude and the decay time.  \cite{wu2015}
clearly demonstrate (see their Fig.~5) that the flares with the largest energy releases are those with the longest
decay times. The same is also true when X-ray flares are considered, some examples are presented by
\cite{tsuboi2000} for energetic flares on the protostars YLW15 and \cite{schmitt1999} and \cite{schmitt2003}
present examples for flares on Algol. 

Fortunately, in the case of the AB~Dor flare on Jan  7$^{}$,
we have two eROSITA data points in the decay phase, which allows us to put rather stringent constraints on the
overall energy output.  In the idealistic case of a purely exponentially decaying flare with an instantaneous rise phase, 
the decay time can be easily computed from
the measured amplitudes and times (A$_1$, t$_1$) and (A$_2$,t$_2$) through $t_{\mathrm{decay}}$ = $\frac{t_2 - t_1}{ln(A_1/A_2)}$,
which in our case yields a decay time value of about 10500~seconds, with some error.  
The amplitude A$_0$ at flare onset
is then given by A$_1$ $\times$ $e^{\frac{\Delta t}{t_{\mathrm{decay}}}}$, where $\Delta t$ denotes the (unknown) time
difference between flare peak and the first available measurement.   We stress that our ignorance of  $\Delta t$ constitutes
the major source of error in our estimates.   However, we can constrain $\Delta t$:   experience from other flares
leads us to expect that the flare peak should have occurred a short
time after the optical flare peak, which suggests $\Delta t$  $\approx$ 7000~s. We know for sure that during
the scan prior to the optical flare peak no X-ray flare was ongoing (see Fig.~\ref{abdor_lc3}), that is, we have
$\Delta t$  $<$ 14000~s.   We therefore expect A$_0$ $\approx$ 2 $\times$ A$_1$, and we know for sure
 A$_0$ $<$ 4 $\times$ A$_1$.   Clearly, $\Delta t$  could be smaller than 7000~s, which would then reduce the total
 energy output in X-rays.   A purely exponential decay is admittedly only an approximation, and
 the rise phase is missing altogether from our considerations.  In our case, there may be some substructure in the optical rise phase,
 while the decay phase appears very smooth. We therefore argue that it is likely that the X-ray decay was also
 very close to being exponential, and in practically all major TESS flares the rise phases are much shorter than the decay phases. 
 Considering all of this evidence, we find it difficult to avoid the conclusion that the
 X-ray energy release in the event from Jan 7$^{}$ was significantly less than the optical energy release.
 
 \subsection{The bigger picture}
 
 There is broad agreement in the literature that the primary energy release of solar and stellar flares takes place in the corona and involves
magnetic energy and nonthermal energetic particles.   The nonthermal particles produced in the corona propagate along the magnetic
field lines to reach denser and cooler regions, where they dissipate their energy, thus producing heating.  
As a consequence, both the optical emission recorded by TESS as well as the soft X-ray emission
recorded by eROSITA are secondary if not tertiary flare products, and therefore an assessment of the overall energetics and
the contribution of the various sources and wavelength ranges to the overall energy budget is notoriously
difficult even for solar flares.  For the case of the large solar X17 flare from October 25,$^{}$ 2003, \cite{woods2004}
present such an analysis and conclude that  the majority of the energy of the flare is
released at wavelengths longward of 200~nm and shortward of 27~nm.  Specifically, \cite{woods2004}
find that 19\% of the observed  total solar irradiance measurements (TSI) change occurs in the XUV and X-ray range, while the UV range between
270~\AA \,and 2000~\AA \,contributes only 3.7\%, which means that almost 80\% of the energy output is observed
in the optical.  Thus, the ratio of X-ray to optical output is about 1:4,  while for the large flare on AB~Dor we estimate
1:20 with, admittedly, considerable uncertainty.  

The immediate question that comes to mind is whether this is a general property of AB~Dor flares. The flare
observed on Dec 22 presents an obvious counter example, because its peak X-ray luminosity of 4 $\times$ 10$^{30}$ erg/s is
accompanied by very little or possibly no detectable optical emission.   Nevertheless, we need to bear in
mind the fact that optical flare light curves are affected by projection and limb-darkening
effects.  A very nice example in this context is provided by the Sun, as already discussed by
\cite{schmitt2019}:  the already mentioned  solar flare from October 25$^{}$ 2003 is the largest one
so far observed in TSI, for which \cite{kopp2005} present a detailed analysis and
derive a total flare energy release of 5~$\times$~10$^{32}$erg.  However, probably the largest
solar flare measured in modern times, namely the class X40 flare discussed by \cite{brodrick2005}
occurring on November $3^{}$, 2003,  right at the solar limb in the very same active region, was 
consequently ---in contrast to the X-ray range---  only very faintly visible in TSI data \citep{kopp2005}.   
Therefore, the {observed} ratio of the X-ray and optical outputs may differ a lot, while the {intrinsic}
ratio may not.  In the solar case, the flare site is usually known and projection and limb-darkening
effects can be assessed, while for spatially unresolved stars the flare site on the stellar surface remains typically unknown
in purely photometric data. On the other hand, as shown by \cite{wolter2008}, in the case of the active
star BO~Mic, a star actually quite similar to AB~Dor, Doppler imaging does in fact allow 
the flare site to be located on the star.

\section{Conclusions}

We present the eROSITA observations of the nearby ultra-active star AB~Dor from the first three
eROSITA surveys; the third (eRASS3) survey was accompanied by simultaneous TESS observations
in the optical.  We find a ``base'' level of AB~Dor's X-ray flux, which appears essentially constant over 1.5 years,
and no evidence for rotational modulation in the X-ray flux, which we attribute to the
saturated nature of AB~Dor's corona.   During the eRASS3 observation, a major flare occurred on
AB~Dor, and the large amount of energy released makes it the largest ever observed on  this star.  Estimating
the total X-ray output from the simultaneously measured eROSITA light curve, we find an X-ray output at least
a factor of ten smaller.   A comparison to large solar flares suggests quite a few similarities and we point
out the limitations of optical observations, which usually do not reveal the flare site on the star.  We therefore
conclude that the measurement of the total energy budget of solar and stellar
flares is a challenge, and it remains to be seen whether big flares in general emit intrinsically more energy in the
optical than in the X-ray range.

\begin{acknowledgements}
This work is based on data from eROSITA, the soft X-ray instrument aboard SRG, a joint
Russian-German science mission supported by the Russian Space Agency
(Roskosmos), in the interests of the Russian Academy of Sciences 
represented by its Space Research Institute (IKI), 
and the Deutsches Zentrum f\"ur Luft- und Raumfahrt (DLR). 
The SRG spacecraft was built by Lavochkin Association (NPOL) 
and its subcontractors, and is operated by NPOL with support from
IKI and the Max Planck Institute for Extraterrestrial Physics (MPE). 
The development and construction of the eROSITA X-ray instrument was
led by MPE, with contributions from the Dr.\ Karl Remeis Observatory 
Bamberg \& ECAP (FAU Erlangen-N\"urnberg), the University of Hamburg 
Observatory, the Leibniz Institute for Astrophysics Potsdam (AIP), 
and the Institute for Astronomy and Astrophysics of the University 
of T\"ubingen, with the support of DLR and the Max Planck Society. 
The Argelander Institute for Astronomy of the University of Bonn and 
the Ludwig Maximilians Universit\"at Munich also participated in the science 
preparation for eROSITA.
The eROSITA data used for this paper were
processed using the eSASS/NRTA software system developed by the
German eROSITA consortium.
This paper includes data collected by the TESS mission, which are publicly available from the 
Mikulski Archive for Space Telescopes (MAST).  

\end{acknowledgements}

\end{document}